\documentclass[a4paper,fleqn,usenatbib]{mnras}

\usepackage{newtxtext,newtxmath}

\usepackage[T1]{fontenc}
\usepackage{ae,aecompl}

\usepackage{graphicx}	
\usepackage{amsmath}	
\usepackage{amssymb}	
\usepackage{subfig}     

\newcommand{\vu}{\mathbf{u}}
\newcommand{\ndvu}{\tilde{\mathbf{u}}}
\newcommand{\ndp}{\tilde{P}}
\newcommand{\ndrho}{\tilde{\rho}}

\title[Dusty fluids]{Dynamics of Dusty Vortices I: Extensions and limitations of the terminal velocity approximation}

\author[F. Lovascio \& S.-J. Paardekooper]{
Francesco Lovascio$^{1}$\thanks{E-mail: f.lovascio@qmul.ac.uk (FL)},
and Sijme-Jan Paardekooper$^{1}$
\\
$^{1}$School of Physics and Astronomy, Queen Mary, University of London, , E1 4NS, UK\\
}

\date{Accepted 22nd July 2019. Received 11th June 2019}

\pubyear{2019}

\begin{document}
\label{firstpage}
\pagerange{\pageref{firstpage}--\pageref{lastpage}}
\maketitle

\begin{abstract}
Motivated by the stability of dust laden vortices, in this paper we study the terminal velocity approximation equations for a gas coupled to a pressureless dust fluid and present a numerical solver for the equations embedded in the FARGO3D hydrodynamics code. We show that for protoplanetary discs it is possible to use the baricenter velocity in the viscous stress tensor, making it trivial to simulate viscous dusty protoplanetary discs with this model. We also show that the terminal velocity model breaks down around shocks, becoming incompatible with the two fluid model it is derived from. Finally we produce a set of test cases for numerical schemes and demonstrate the performance of our code on these tests. Our implementation embedded in FARGO3D using an unconditionally stable explicit integrator is fast, and exhibits the desired second order spatial convergence for smooth problems.
\end{abstract}



\section{Introduction}
Dust is ubiquitous in space. In most astrophysical scenarios dust is coupled to a gaseous phase, such that its evolution is determined by an interplay of gas dynamics, external forces, and self gravity. Gas-dust dynamics plays an important role in AGB star winds \citep[e.g.][]{1989A&A...223..227D, 2018A&ARv..26....1H}, and planetary and substellar atmospheres \citep[e.g.][]{2017A&A...608A..70J, 2008MNRAS.385L.120J, Taylor2007} as well as in protoplanetary discs. The behaviour of dust in such a mixture can be understood as an ensemble of dust particles feeling the external fields, as well as a drag force from the surrounding gas.
Many astrophysical flows, especially protoplanetary discs tend to have very well but not perfectly coupled dust and gas \citep[e.g.][]{2014Dust_Evolution}. Effectively modelling the behaviour of these dusty gases is key to understanding many astrophysical phenomena, from planet formation to the formation of complex molecules in the inter stellar medium \citep[e.g.][]{WAKELAM20171}. This is especially true when tying the models to infrared (IR) observations from telescopes like the Atacama Large Millimeter Array (ALMA) as observed IR emission from gas clouds and discs is largely thermal emission from millimeter and sub millimeter dust particles in these objects. The brightness of dust continuum emission makes it in general easier to observe than gas. This often makes dust the target for observations of protoplanetary discs, \citep[see e.g][]{10.1093/mnras/sty2653}.

Dust comprises only a small fraction of interstellar matter (\textasciitilde1\% by mass), but its effects on the dynamics of flows can be important such as with the damping of vortices \citep{dvortices2D}. In other cases the dynamics of the dust phase in a gas flow can be quite different than the underlying gas flow. For example in the case of protoplanetary discs it is much easier to clear gaps in the dust than it is in the gas, leading to much deeper gaps in the dust density than in the gas density \citep[e.g.][]{2004A&A...425L...9P}. For protoplanetary discs, dust is both important in the evolution of the disc, among other things playing a fundamental role in planet formation, and key to observations. The highest resolution data for protoplanetary discs has been produced by ALMA \citep[e.g.][]{2016ApJ...820L..40A} taking observations in sub-millimeter continuum emission. This fine dust is in the very well, but not perfectly coupled regime. This makes understanding the behaviour of well coupled dust in a gas crucial also in the interpretation of observations for protoplanetary discs. 

 One area where dust-gas interactions can play an important role is in the evolution of dust-laden vortices. These dusty vortices are understood to be important in the dynamics of protoplanetary discs. In gas only discs, vortices arising from instabilities may be long lived \citep[e.g.][]{2014ApJ...788L..41F}, or even self sustaining \citep[e.g.][]{2010A&A...513A..60L}. Protoplanetary discs however, contain dust, which has been shown to collect in vortices \citep{BargeSommeria} and may therefore aid the process of planet formation \citep{LyraJohansen}. These dusty vortices do not behave like their gas only counterparts, with dust damping the vortex and driving it unstable as shown in \cite{2014ApJ...795L..39F}, \cite{Chang_2010} and \cite{2014MNRAS.445.4409R}. 
 
 To fully capture the physics of the gas-dust mixture, a model that solves the motion of the individual dust particles in the gas flow would be required, studying this kind of model though is currently not computationally viable for large systems, and is virtually intractable analytically. Further study of the driving mechanisms behind these instabilities requires high resolution simulations as well as analytical study of the governing equations; both are good reasons to apply simplified models to the problem.

Several approaches have been taken to model dusty astrophysical flows. The cheapest methods use a small number of test particles feeling a drag force due to the fluid around them and an additional force due to any external potential field they may be in, like was done by \cite{2018arXiv181112841M} or \cite{10.1093/mnras/stv1486}. It is more challenging to simulate the back reaction of the dust on the gas, many methods attempting to do this also model the dust as a fluid. This kind of approximation is possible when the dust particles' mean free path through the fluid is short compared to the typical length scale of the flow, allowing for the description of the dust through bulk properties \citep{Garaud_2004}. Examples of this kind of model are the Lagrangian dust model described in \cite{Osiptsov2000} or two fluid Eulerian models like \cite{2004A&A...417..361J}. This kind of problem can also be quite intuitively approached using an SPH (smoothed particle hydrodynamics) approach, where some of the fluid particles are dust, and others are gas, like is done in \cite{10.1093/mnras/stv996}.

As a possible further simplification to modelling back reacting dust in a gas, one may consider one fluid terminal velocity approximations like the ones from \cite{LaibePriceDust} and \cite{LinYoudin2017}. These approximations require for less equations to be solved, while still capturing the effects of the dust on the gas. With this kind of approximation it becomes possible to more easily run simulations at resolutions where with two fluids it would have required large amounts of computing power, or allow for parameter studies to be done where with two fluids only a single run could have been done. This is especially true for very well coupled dust gas mixtures, as the two fluid equations become stiffer with stronger coupling between the dust and gas, while the opposite is true for the terminal velocity approximation. These one fluid approximations also have the added benefit of being more tractable with analytic methods.
As one fluid terminal velocity approximations have become more popular, several numerical codes have been written to solve this kind of problem like in \cite{LinYoudin2017}, \cite{2018MNRAS.476.2186H} for multiple grain sizes, or our own variation on FARGO3D described in section \ref{sec:methodFargo}. 

Motivated by our interest in the stability of dusty vortices, in this paper we try to extend understanding of the terminal velocity approximation, formulating a viscous version of the equations, and study the limitations and breakdown of the approximation around sharp pressure gradients and shocks. We also propose some analytic solutions to possibly be used as test cases for one fluid terminal velocity approximation solvers, similarly to what was recently done by \cite{tw0_fluid_shocks_wardle} for two fluid dusty gas hydrodynamics solvers. 

The rest of this paper is structured as follows. In section \ref{sec:Visc} we re-derive the terminal velocity approximation, showing how it can consistently be extended to viscous discs without any additional source terms. We then in section \ref{sec:Tests} detail several analytic solutions of the equations, a small amplitude wave solution, a solution for shock profiles, and a purely diffusive solution. In section \ref{sec:Prob} we compare the one fluid and two fluid shock solutions and show that a first order error arises in the terminal velocity approximation around shocks. In section \ref{sec:methodFargo} we detail an implementation embedded in FARGO3D and its performance. In section \ref{sec:Res} we show the performance of our implementation of FARGO3D on the analytical test problems derived in section \ref{sec:Tests}. In sections \ref{sec:Dis} and \ref{sec:Conc}, we discuss and summarise the results of the paper.

\section{Dusty Gases}\label{sec:Visc}
A dusty fluid flow can be modelled as two interacting fluids, with a gas phase and a pressureless cold dust phase. The description of dust as a fluid is possible provided that the dust is coupled to the gas, allowing the dust particles to exhibit collective behaviour \citep{Garaud_2004}. A gas and cold dust phase can be understood as only coupled by a drag force. The drag force is modelled as an Epstein drag. This is applicable high Knudsen number (the dimensionless relation between characteristic length and mean free path in a system) regime. This means that the dust particles are smaller than the gas molecules' mean free path. The two fluid model however is complex to deal with, having at least two additional fields and equations than a pure gas in the form of the continuity and momentum equation for the dust, which require the addition of a dust density and dust velocity field. This makes the problem both computationally expensive to solve numerically, and difficult to study analytically. Several approaches have been taken to reduce the complexity of the problem, the one we are mainly interested in is the terminal velocity approximation. The terminal velocity approximation is a first order in Stokes number (a dimensionless number describing the relation between drag and momentum of a body in a flow) approximation on the one fluid formulation of the dusty gas equations and hence works best for small particles. 
Here we run through the derivation of the one fluid approximation for a pressureless dust fluid coupled to an isothermal gas. We use some arguments from dimensional analysis to demonstrate how the equations can be extended to include viscous terms in the case of  protoplanetary discs, at no additional cost.   
\subsection{From two fluids to one fluid approximation}
The two fluid dust equations describe a dusty fluid in a potential field $\Phi$ as two fluids coupled by a drag force parametrised as a stopping time $t_s$. One of the fluids is the pressure-less ''dust fluid'' with velocity $\vu_d$ and density $\rho_d$, the other is the isothermal gas described by its velocity $\vu_g$, density $\rho_g$, and a sound speed $c_s$, the pressure in this fluid is given by $P=c_s^2 \rho_g$. For dust coupled to a viscous gas we also add the stress tensor $\mathsf{T}_g$ in the gas equations.
	\begin{align}
	&\begin{aligned}
	\label{eq:twofluid.1}
		&\partial_t\rho_g+\nabla\cdot\left(\rho_g \vu_g\right) =0 
	\end{aligned}\\
	&\begin{aligned}
	\label{eq:twofluid.2}
        &\partial_t \left(\rho_g \vu_g\right)+\nabla\cdot\left(\rho_g \vu_g \vu_g\right)+c_s^2 \nabla\rho_g =\nabla\cdot \mathsf{T}_g -\rho_g\nabla\Phi -\\
        & \frac{\rho_d \rho_g}{t_s\rho}\left(\vu_g-\vu_d\right)
        \end{aligned}\\
    &\begin{aligned}
    \label{eq:twofluid.3}
        &\partial_t \rho_d+\nabla\cdot\left(\rho_d \vu_d\right) =0
    \end{aligned}\\
    &\begin{aligned}
    \label{eq:twofluid.4}
         &\partial_t \left(\rho_d \vu_d\right)+\nabla\cdot\left(\rho_d \vu_d \vu_d\right) = -\rho_g\nabla\Phi -\frac{\rho_d\rho_g}{t_s\rho}\left(\vu_d-\vu_g\right)
    \end{aligned}
	\end{align}
The velocity can then be re-expressed as a bulk velocity $\vu$ such that: $\vu=(\rho_g \vu_g +\rho_d \vu_d)/\rho$ where $\rho=\rho_d +\rho_g$, the total density, and the velocity difference between gas and dust $\Delta \vu = \vu_d - \vu_g$. To describe the gas to dust ratio we choose a gas to total mass ratio $f_0=\rho_g/\rho$, for an isothermal gas $f_0=P/c_s^2\rho$, as the pressure $P=c_s^2 \rho_g$. This allows for the reconstruction of the two fluid equations as a set of one fluid equations without need for any approximations.  
\begin{align}
	&\begin{aligned}
	\label{eq:OnefluidNoTerminal.1}
		&\partial_t\rho+\nabla\cdot\left(\rho \vu\right) =0 
	\end{aligned}\\ 
	&\begin{aligned}
	\label{eq:OnefluidNoTerminal.2}
        &\partial_t \left(\rho \vu\right)+\nabla\cdot\left(\rho \vu \vu + \frac{P}{c_s^2} \left(1-\frac{P}{c_s^2\rho}\right)\Delta \vu \Delta \vu \right)+\nabla P =\\
        &\nabla\cdot \mathsf{T_g} -\rho\nabla\Phi  
    \end{aligned}\\  
    &\begin{aligned}
	\label{eq:OnefluidNoTerminal.3}
        &\partial_t P+\nabla\cdot\left(P\vu-P\left(1-\frac{P}{c_s^2\rho}\right)\Delta \vu\right) =0
    \end{aligned}\\   
    &\begin{aligned}
	\label{eq:OnefluidNoTerminal.4}
         &\partial_t \Delta \vu +\nabla\cdot\left(\vu\Delta \vu - \left(\frac{1}{2}-\frac{P}{c_s^2\rho}\right)\Delta \vu \Delta \vu\right) = \\
         &-\frac{\Delta \vu}{t_s} + c_s^2 \frac{\nabla P}{P}- c_s^2\frac{\nabla\cdot\mathsf{T_g} }{P}. 
    \end{aligned}
\end{align}
The non dimensional formulation of the 1 fluid gas dust equations is constructed by choosing a typical length scale $L$, velocity $\bar{u}$, density $\bar{\rho}$ and gas fraction $\bar{f_0}=\bar{P}/(c_s^2 \bar{\rho})$, and decomposing the flow fields, $\xi = \bar{\xi}\tilde{\xi}$ where $\tilde{\xi}$ is the dimensionless field, 
\begin{align} 
	&\begin{aligned} 
	\label{eq:twofluidnondim.1}
		&\partial_t\ndrho+\nabla\cdot\left(\ndrho \ndvu\right) =0 
    \end{aligned}\\
    &\begin{aligned}
    \label{eq:twofluidnondim.2}
        &\partial_t \left(\ndrho \ndvu\right)+\nabla\cdot\left(\ndrho \ndvu \ndvu + \ndp \bar{f_0} \left(1-\frac{\ndp\bar{f_0}} {\ndrho}\right)\Delta \ndvu \Delta \ndvu \right)+\frac{\bar{f_0}}{\mathrm{Ma^2}}\nabla \ndp =\\
        &\frac{\bar{f_0}\nabla\cdot \tilde{\mathsf{T}}_g}{\text{Re}} -\frac{\ndrho\nabla\Phi}{\mathrm{Fr^2}} \\ 
    \end{aligned}\\
    &\begin{aligned}
    \label{eq:twofluidnondim.3}
        &\partial_t \ndp+\nabla\cdot\left(\ndp\ndvu-\ndp\left(1-\frac{\ndp\bar{f_0}}{\ndrho}\right)\Delta \ndvu\right) =0  
    \end{aligned}\\
    &\begin{aligned}
    \label{eq:twofluidnondim.4}
        &\partial_t \Delta \ndvu +\nabla\cdot\left(\ndvu\Delta \ndvu - \left(\frac{1}{2}-\frac{\ndp\bar{f_0}}{\ndrho}\right)\Delta \ndvu \Delta \ndvu\right) = \\
        &-\frac{\Delta \ndvu}{\tilde{t_s} \text{St}} + \frac{\nabla \ndp}{\mathrm{Ma^2}\ndp}-\frac{\nabla\cdot\tilde{\mathsf{T}}_g }{\ndp\text{Re}}.
    \end{aligned}\\
    \notag
\end{align}
 The dimensionless numbers produced are the Froude number $\mathrm{Fr}=\bar{u}/\sqrt{gL}$, the Stokes number $\mathrm{St}=t_s \bar{u}/L$, the Mach number $\mathrm{Ma}=\bar{u}/c_s$, and the Reynolds number $\mathrm{Re}=\bar{u}L/\bar{\nu}$. We keep a non-dimensional stopping time in the equations for clarity, which is understood to be $\mathcal{O}(1)$. The nondimensional viscous stress tensor $\tilde{\mathsf{T}}_g$ is given by,
\begin{equation}\label{eq:strestensnondim}
\tilde{\mathsf{T}}_g=\tilde{\nu} \ndp \left[\nabla \ndvu_g + \left(\nabla \ndvu_g\right)^T -\frac{2}{3}\left(\nabla \cdot \ndvu_g\right)\mathrm{I} \right],
\end{equation}
where we keep a viscosity $\tilde{\nu}$ which is understood to be $\mathcal{O}(1)$.

\subsection{Asymptotic expansions and the $\mathrm{St}\ll 1$ approximation}
Using a power series expansion in $\mathrm{St}$ for the flow fields such that a flow variable $\xi(x,t,\mathrm{St})$ becomes
\begin{equation}\label{eq:asymexp}
\xi(x,t,\mathrm{St})=\xi^{(0)}(x,t)+\mathrm{St}\xi^{(1)} (x,t)+ \mathcal{O}(\mathrm{St^2}),
\end{equation}
we thus exploit the assumption that we are in the $\mathrm{St}\ll 1$ regime. The only flow field that depends explicitly on $\mathrm{St}$ is $\Delta \ndvu$, since equation (\ref{eq:twofluidnondim.4}) is the only one with a term of $\mathrm{St}$. As the coupling approaches perfect coupling we expect the velocity difference between the dust and gas phases to go to zero so, 
\begin{equation}
    \lim_{\mathrm{St}\to 0}(\Delta \ndvu) = 0,
\end{equation}
which implies that $\Delta \ndvu^{(0)}=0$.
Substituting the expansion for $\Delta\ndvu$ into equation (\ref{eq:twofluidnondim.4}), we obtain,
\begin{equation}
    \Delta \ndvu^{(1)}=\frac{\tilde{t_s}}{\ndp}\left(\frac{\nabla \ndp}{\mathrm{Ma^2}}-\frac{\nabla\cdot\tilde{\mathsf{T}}_g }{\mathrm{Re}}\right)+\mathcal{O}(\mathrm{St})
\end{equation}
this expression can be substituted back into the expansion of $\Delta \ndvu$ to give the expression
\begin{equation} \label{eq:RelativeMagnitudes}
    \Delta \ndvu = \mathrm{St}\frac{\tilde{t_s}}{\tilde{P}}\left(\frac{\nabla \ndp}{\mathrm{Ma^2}}-\frac{\nabla\cdot\tilde{\mathsf{T}}_g }{\mathrm{Re}}\right) +\mathcal{O}(\mathrm{St}^2).
\end{equation}
The form of the viscous terminal velocity approximation depends on the relative magnitudes of Re, St and Ma. For it to be possible to extract a simple terminal velocity approximation from these equations we have to be in a regime where $\mathrm{Re}\gg\mathrm{Ma}^2$ which can be formalised by setting $\mathrm{Re}=\mathcal{O}\left(\mathrm{St}^{-1}\right)$.
The viscous term in (\ref{eq:RelativeMagnitudes}) then gets pushed to higher order and taking this first order approximation in $\mathrm{St}$ for an inviscid gas one then retrieves a terminal velocity approximation as derived by \cite{LinYoudin2017}. 

Having taken the terminal velocity approximation, (\ref{eq:twofluidnondim.4}) vanishes, so the viscous stress tensor $\mathrm{T}_g$ (\ref{eq:strestensnondim}) only appears in equation (\ref{eq:twofluidnondim.2}), with factor of  $\mathrm{Re^{-1}}$. As shown in (\ref{eq:RelativeMagnitudes}), for the terminal velocity approximation to hold $\mathrm{Re}$ must be larger than $\mathcal{O}\left(\mathrm{St}^{-1}\right)$, so
\begin{align}
    &\mathrm{Re^{-1}}=\mathcal{O}(\mathrm{St}),\\
    &\ndvu_g=\ndvu+\mathcal{O}(\mathrm{St});
\end{align} 
such that
\begin{equation}
\frac{\ndvu_g}{\mathrm{Re}}=\frac{\ndvu}{\mathrm{Re}}+\mathcal{O}\left(\frac{\mathrm{St}}{\mathrm{Re}}\right)=\frac{\ndvu}{\mathrm{Re}}+\mathcal{O}\left(\mathrm{St^2}\right),
\end{equation}
the stress tensor can be rewritten as
\begin{equation}
\begin{aligned}
\label{eq:appstress}
&\frac{\tilde{\mathsf{T}}_g}{\mathrm{Re}} =\frac{1}{\mathrm{Re}}\tilde{\nu} P \left[\nabla \ndvu + \left(\nabla \ndvu\right)^T -\frac{2}{3}\left(\nabla \cdot \ndvu\right)\mathrm{I} \right] +\mathcal{O}(\mathrm{St^2})\\
&=\frac{\tilde{\mathsf{T}}}{\mathrm{Re}}+\mathcal{O}(\mathrm{St^2}).
\end{aligned}
\end{equation}
It can be shown that in fact, for protoplanetary discs $\mathrm{Re}\gg\mathrm{Ma^2}$ for $\alpha$ viscosity. Considering a protoplanetary disc with Reynolds number $\mathrm{Re}={\bar{u}\bar{L}}/{\bar{\nu}}$ one can take the characteristic velocity $\bar{u}$ to be the Keplerian velocity $u_K=c_s\mathrm{Ma}$, such that $\mathrm{Re}=c_s \mathrm{Ma} \bar{L}/\bar{\nu}$. In the alpha viscosity formalism $\nu = \alpha c_s  H$ \citep{ShakSun}, taking $\bar{L}$ as the orbital radius makes $\bar{L}/H=\mathrm{Ma}$ so the Reynolds number becomes $\mathrm{Re}=\mathrm{Ma^2}/\alpha$, given that $\alpha \ll 1$ \citep{2017ApJ...837..163R}; $\mathrm{Re}\gg\mathrm{Ma^2}$.
It is therefore correct to first order in $\mathrm{St}$ to use the gas viscous stress tensor for protoplanetary discs. The nondimensional equations can therefore be rewritten in one fluid form to include a viscous term as well, 

	\begin{align}
		&\partial_t\ndrho+\nabla\cdot \left(\ndrho \ndvu\right) =0 \label{eq:onefluidnondim.1}\\ 
        &\partial_t \left(\ndrho \ndvu\right)+\nabla\cdot\left(\ndrho \ndvu \ndvu \right)+\frac{\bar{f_0}}{\mathrm{Ma^2}}\nabla \ndp =\frac{\bar{f_0}\nabla\cdot \tilde{\mathsf{T}}}{\text{Re}} -\frac{\ndrho\nabla\Phi}{\mathrm{Fr^2}}  \label{eq:onefluidnondim.2}\\  
        &\partial_t \ndp+\nabla\cdot\left(\ndp\ndvu\right) =\frac{\mathrm{St}}{\mathrm{Ma^2}}\nabla\cdot\left[\tilde{t_s} \left(1-\frac{\ndp}{c_s^2\ndrho}\right)\nabla \ndp\right].\label{eq:onefluidnondim.3}\\ 
        \notag
    \end{align}
    In dimensional form these become,
    \begin{align}
		&\partial_t\rho+\nabla\cdot \left(\rho \vu\right) =0 \label{eq:onefluiddim.1}\\
        &\partial_t \left(\rho \vu\right)+\nabla\cdot\left(\rho \vu \vu \right)+\nabla P =\nabla\cdot \mathsf{T} -\rho\nabla\Phi  \label{eq:onefluiddim.2}\\  
        &\partial_t P+\nabla\cdot\left(P\vu\right) =\nabla\cdot\left[t_s \left(1-\frac{P}{c_s^2\rho}\right)\nabla P\right].\label{eq:onefluiddim.3}\\ 
        \notag
    \end{align}
This is the same set of equations derived in \cite{LinYoudin2017}, the locally isothermal terminal velocity approximation (LITVA), with the addition of a viscous term. The equations (\ref{eq:onefluiddim.1}-\ref{eq:onefluiddim.3}) can also be obtained through assuming an isothermal gas in the \cite{LaibePriceDust} terminal velocity approximation. The dust then behaves as a cooling term acting on the gas, removing the need to solve a separate dust advection equation. 
\section{Test Cases}\label{sec:Tests}
To test our implementation of a one fluid gas dust solver we consider some analytic test cases. We produced analytic solutions for three initial conditions, a small amplitude wave, a steady state shock, and a pure diffusion problem. The latter two could be especially useful as they complement the existing studies of analytic solutions already found in literature like two fluid shock solutions \citep{tw0_fluid_shocks_wardle} and a dusty wave dispersion analogous to our first test case \citep{2011MNRAS.418.1491L}. Analytic solutions like these can not only be used as test cases but also give some insight to the behaviour of the equations and better understand simulation results.
\subsection{Dusty Wave Solutions}\label{sec:Tests:wave}
Using linear perturbations it is possible to obtain a dispersion relation for small amplitude sound waves in a dusty gas similarly to what \cite{2016ascl.soft02004L} did for their one fluid formulation. Starting from  equations (\ref{eq:onefluiddim.1}-\ref{eq:onefluiddim.3}) in the inviscid regime and linearising by applying a small wavelike perturbation $\delta(F) = F'\psi$ to a background $F_0$, where $\psi = e^{i(kx+\omega t)}$ and $F' \ll F_{0}$. Given that $F'\ll F_{0}$ terms which are of order $\mathcal{O}(F'^2)$ can be neglected. This leaves us with
\begin{equation}
	\begin{aligned}\label{eq:linearised}
		&i\omega \delta\left(\rho\right) + ik\rho_0 \delta\left(\vu\right)=0 \\
        &i\omega \rho_0 \delta\left(\vu\right)=ik\delta\left(P\right)\\
        &i\omega \delta\left(P\right)+ikP_0\delta \left(\vu\right) = -c_s^2t_s\left(1-\frac{P_0}{c_s^2 \rho_0}\right)k^2\delta\left(P\right)
	\end{aligned}
\end{equation}from which a dispersion relation can be calculated. This dispersion relation is complex, as there are both wavelike and damping like parts to the solution. 

\begin{equation}
   \omega=\frac{ik^2D_c}{2} \pm |k|\sqrt{\frac{P_0}{\rho_0}-\frac{D_c^2}{4}}, 
\end{equation}
where $D_c=c_s^1 t_s \left(1-P_0/c_s^2 \rho_0 \right)$ or,
\begin{equation}\label{eq:disp}
	\omega=\frac{ic_s^2t_sf_dk^2 \pm \sqrt{-c_s^4t_s^2f_d^2k^4+4k^2c_s^2(1-f_d)}}{2}.
\end{equation}
Having a complex dispersion relation requires the perturbations $F'$ to be complex too, with the consequence that for dusty sound waves the wave in pressure lags the wave in density by a phase proportional to $f_d$, the gas to dust ratio, defined as $f_d = 1- P/c_s^2 \rho$.
This relation provides an analytic solution of the gas dust equations in the linear regime, which we used to test the convergence properties of our scheme. It should be noted that this solution is a wave damping in time where each branch corresponds to a left travelling or right travelling wave, the superposition of the two producing a damped standing wave. It is important to note that the damping in this solution is entirely due to the dust cooling term and not some viscous term.
\subsection{Dusty Shock Solutions}\label{sec:Test:shock}
\begin{figure}
\includegraphics[width=0.9\columnwidth]{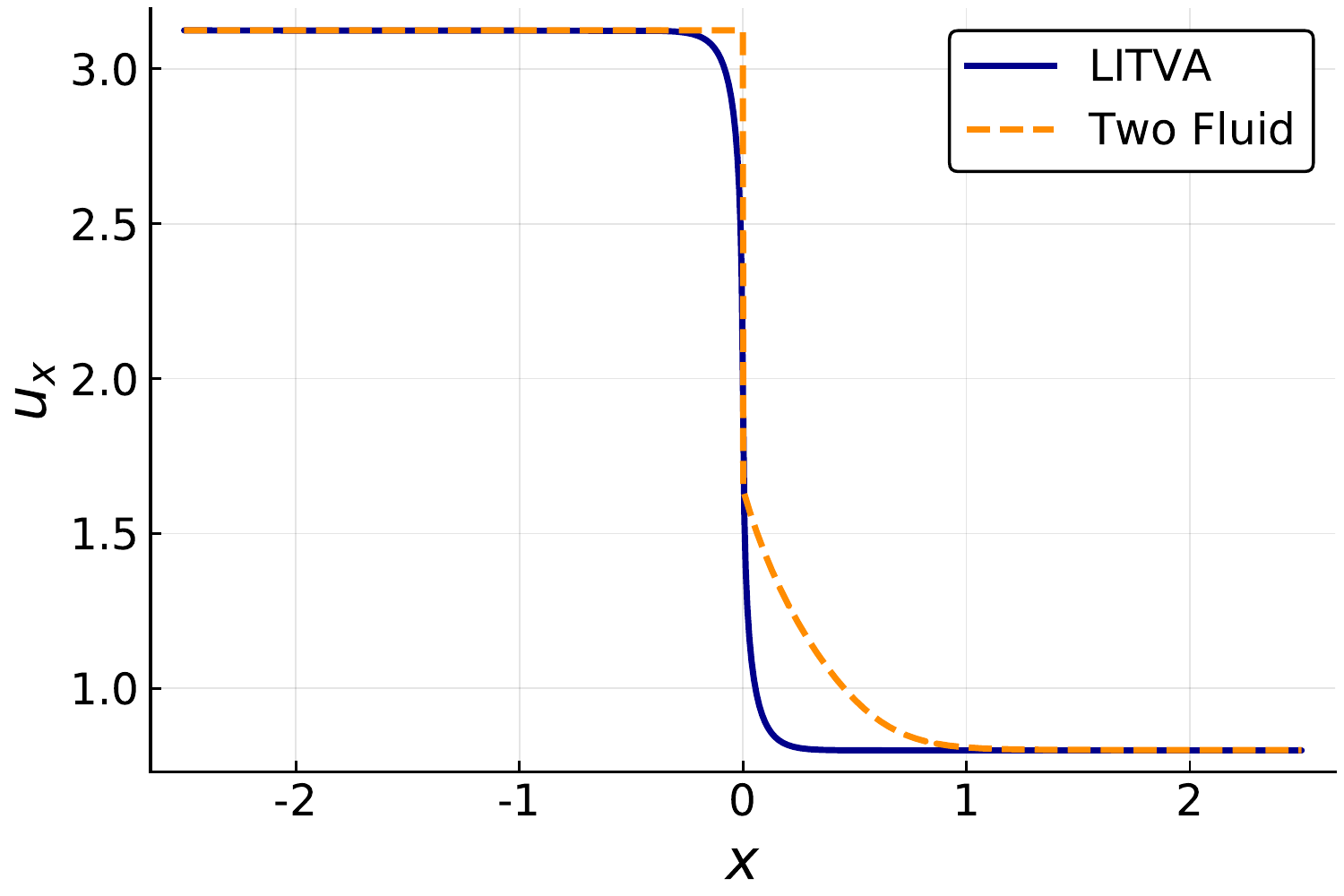}
\caption{A stationary dusty shock profile calculated analytically, overlaid with the two fluid analytic solution, The two quantities overlayed are the barycenter velocity in the two fluid model and in the locally isothermal terminal velocity approximation, the solution shown are for $t_s=0.1$ and $c_s=2.0$. The discrepancy between the two models shown will be discussed in depth in section \ref{sec:Prob}}\label{fig:Shock}
\end{figure}

A solution involving steep gradients and large amplitudes to the one fluid dust gas approximation is the solution around a shock front. While there exists an analytic solution for this case, it should in fact be noted that the terminal velocity approximation breaks down around shocks. This solution is unphysical, as will be discussed more rigorously in section \ref{sec:Prob}. 
Here we construct a steady state shock solution; this solution holds true only when the shock has reached steady state as $t \rightarrow \infty$, and the diffusion term is balanced by the advective terms. We could not construct a full solution like the \cite{SOD19781} solutions, as the diffusive term breaks scale invariance for the system of equations, making a solution especially for the rarefaction wave very hard to find. Starting with the system of equation (\ref{eq:onefluiddim.1}-\ref{eq:onefluiddim.3}) in the inviscid regime, constructed in 1D 
\begin{align}
    &\partial_t(\rho) +\partial_x(\rho u)=0 \label{eq:1D.1}\\
    &\partial_t(\rho u) +\partial_x(\rho u^2 +P)=0 \label{eq:1D.2}\\
    &\partial_t(P) +\partial_x(Pu)=C({P,\rho}) \label{eq:1D.3}\\
    &C({P,\rho})=c_s^2\partial _x \left[t_s\left(1-\frac{P}{c_s^2\rho}\right)\partial_x P\right].\label{eq:1D.4}
\end{align}
Considering the equations(\ref{eq:1D.1}-\ref{eq:1D.3}) in the frame co-moving with the shock and allowing for a steady state to be reached,

    \begin{align}
        &\partial_x(\rho u)=0\label{eq:shocks}\\
        &\partial_x(\rho u^2 +P)=0\label{eq:shocks.2}\\
        &\partial_x(Pu)=C({P,\rho})\label{eq:shocks.3}
    \end{align}
the invariants for these equations (\ref{eq:shocks}) are,
    \begin{align}
	    B=&\rho_R u_R =\rho_Lu_L \label{eq:reimanInv.1}\\
	    D=&\rho_R u^2_R +P_R =\rho_Lu_L^2 +P_L\label{eq:reimanInv.2},
	\end{align}
where $\rho_L,P_L,u_L$ are the "left" preshock values, and $\rho_R,P_R,u_R$ are the "right" postshock values, in a shock travelling from right to left. Additionally $f_{dR} = f_{dL}$,  however $f_d$ is not constant over the shock, as that would require $f_d=0$ or $t_s=0$, i.e. either the no dust or perfect coupling solutions respectively.
The jump conditions determining $u_L$ and $u_R$ can be calculated the same way as they would for the gas only case. The jump depends only on the Mach number of the shock and the soundspeed,
\begin{equation}
    u_R=\frac{c_s'^2}{u_L}.
\end{equation}
In the dusty case however, the soundspeed in the dusty fluid is the modified soundspeed $c_s'=f_0 c_s$
The system of equations can be thus reduced to a single ODE (\ref{eq:sprofile}), which can then be solved numerically to produce the profile of a steady state dusty shock.
\begin{equation}\label{eq:sprofile}
	\frac{P(B-P) - P_L(B-P_L)}{D}=t_s\left(c_s^2 -\frac{P(B-P)}{D^2}\right)\frac{dP}{dx}
\end{equation}
The ODE (\ref{eq:sprofile}) describing the pressure profile of a shock is first order and therefore has some form of analytic solution (even if not necessarily integrable, as is the case here), which can be found through substitution to be,
\begin{equation}\label{eq:anshock}
	\frac{\tanh\left[ \xi \frac{P_c-P}{\rho_Lu_L^2}\right]-\frac{z-z_c}{1-zz_c}}{1-\tanh\left[ \xi \frac{P_c-P}{\rho_Lu_L^2}\right]\frac{z-z_c}{1-zz_c}} = \tanh\left( \frac{\xi x}{u_Lt_s}\right)
\end{equation}
where $\xi=0.5\left(\frac{u_L^2-a^2}{c^2 -a^2}\right)$ and $z=\frac{B-2P}{B-2P_L}$, for shocks $u_L<a<c)$. $P_c$ is the central value of $P$ at the shock; to obtain this value we have to assume that the shock is symmetric, which from the $\tanh(x)$ like solution is a fair assumption (Fig. \ref{fig:Shock}). Equation (\ref{eq:anshock}) can be simplified to 
\begin{equation}
\label{eq:ShockSimp}
\frac{u_0^2 -a_0^2}{2u^2_0}z-\frac{1}{\xi}\tanh^{-1}z = \frac{x}{u_0 t_s}.
\end{equation}
This  expression also has no explicit roots for $z$ and therefore $P_x$. Thus, a numerical root finder like Newton-Rhapson must be used, producing the same results as those obtained through the numerical integration of equation (\ref{eq:shocks}).  For all $\xi <1$ this solution is smooth, however on a large scale, this still can be approximated as a jump described by the perfectly coupled jump conditions. As mentioned before, this solution is only an asymptotic solution, that the fluid relaxes into. The rate at which this solution is approached does not behave intuitively; the diffusion timescale actually grows with $t_s$. It can be shown that this is the case by changing the time coordinate to a ''diffusion time'' $\tau=t\cdot t_s$, which removes $t_s$ from the cooling term, given that $\tau\propto t_s$ the rate at which a time dependant solution approaches the stationary solution; larger stopping times lead to solutions which which evolve slower towards the steady state solution than smaller stopping times. The most obvious case of this is the $t_s=0$ perfect coupling case described by \cite{LaibePriceDust}, where the steady state solution is reached immediately. As with the wave solution from section \ref{sec:Tests:wave} the shock smoothing is entirely due to dust gas interactions. 
\subsection{A purely diffusive solution, and its relation to the porous medium equation}
It is also possible to test the behaviour of a scheme for just the diffusion problem. This is actually a useful test, as it is not obvious that a diffusion solver would be well behaved for nonlinear diffusion problems such as this one. Assuming that the evolution is entirely dominated by the dust source term from equation (\ref{eq:onefluiddim.3}), the only equation that remains is the anomalous diffusion equation described by the time evolution of the cooling term,
\begin{equation}
\label{ed:diff}
\partial_t P=c_s^2 \nabla\cdot\left[t_s \left(1-\frac{P}{c_s^2 \rho}\right)\nabla P\right].
\end{equation}
There were no real initial conditions in the advection diffusion problem for which this would be the case, so this is probably best thought of as mainly a test problem.  
Equation \ref{ed:diff} is a nonlinear diffusion equation as the diffusion operator can be written in the form $\nabla \cdot \left(D_P\nabla P\right)$, which is in fact a sub diffusion equation as $D_P=D' P^{n}$ where  $n>0$ and $D'$ is constant in $P$ \citep{Vzquez2017TheMT}. Through the transformation $f_d=1-\frac{P}{c_s^2\rho}$, assuming $\rho$ is constant, we can obtain the equation 
\begin{equation}
\label{eq:porusmedium}
\partial_t f_d= c_s^2 t_s\nabla \cdot\left(f_d\nabla f_d\right),
\end{equation}
a porous medium equation of the form
\begin{equation}
\label{eq:generalporous}
\partial_t F=\nabla \cdot \left( F^{n}\nabla F\right) .
\end{equation}
The $n=1$ case for the porous medium equation is also called the Boussinesq equation, which is used to describe the filtration of water underground \citep{boussinesq1903mode}. 
From a physical standpoint it is interesting that one would recover something of the form of the Boussinesq equation from the set of equations (\ref{eq:onefluiddim.1}-\ref{eq:onefluiddim.3}). The Boussinesq equation describes a filtration of a fluid (water) through a fixed matrix of particles (soil), which for a case when the stopping time is quite long compared to the characteristic time of the system, is actually quite similar to a dusty gas. 

There exists a well studied self similar solution to equation (\ref{eq:generalporous}), the Barenblatt profile \cite{barenblatt_1996}, which for the $\mathbb{R}^1, n=1$ case is described by the expression 
\begin{equation}
F(x,t)= \left\{
\begin{array}{ll}
      t^{-\frac{1}{3}}\left(\frac{A}{6} - \frac{x^2}{6t^{\frac{2}{3}}}\right)& \frac{A}{6} - \frac{x^2}{6t^{\frac{2}{3}}}\geq 0 \\
      0 & \frac{A}{6} - \frac{x^2}{6t^{\frac{2}{3}}}<0
\end{array} 
\right.
\end{equation}

\begin{figure}
\includegraphics[width=0.9\columnwidth]{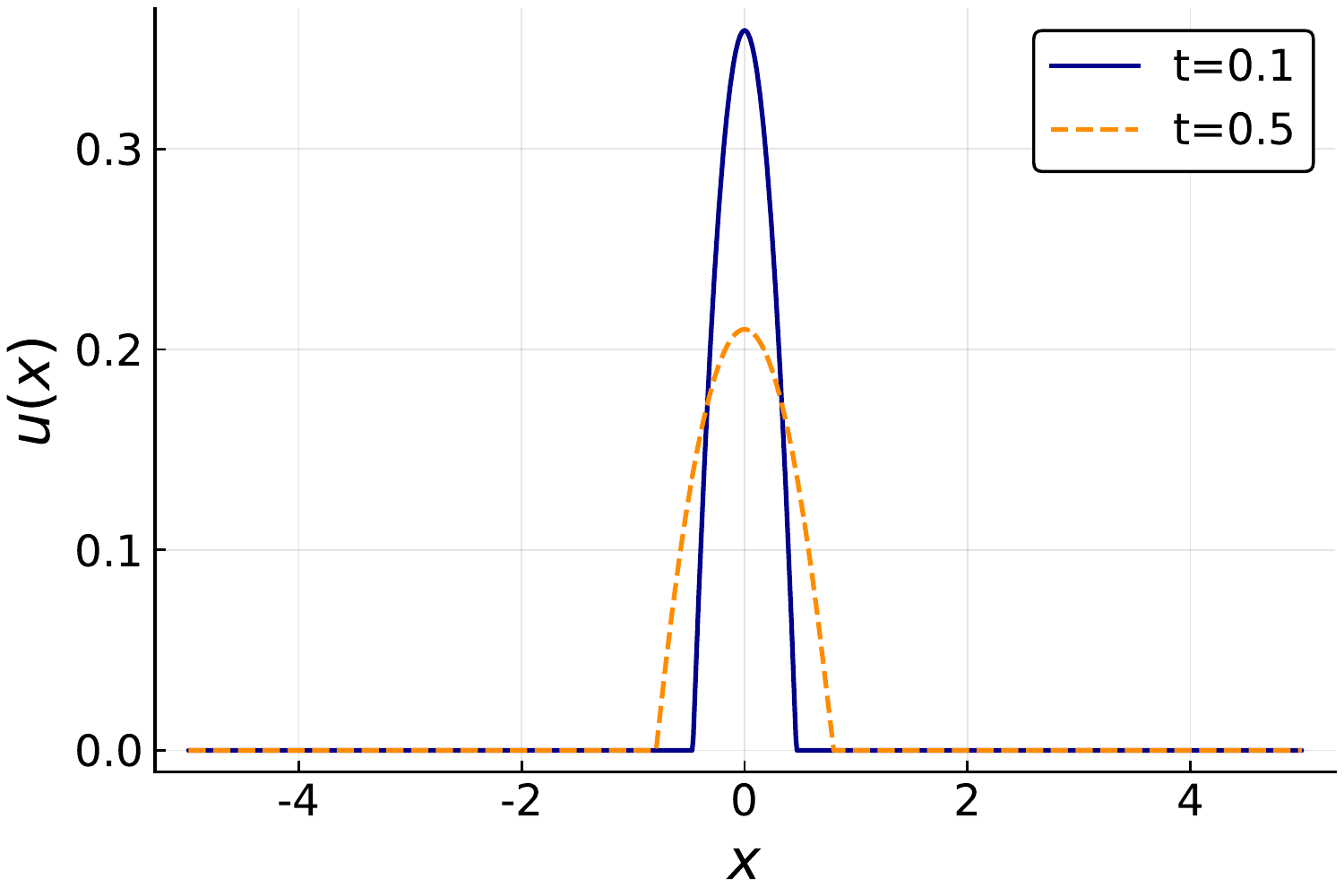}
\caption{The Barenblatt profile at times t=0.1 and t=0.5 for A=1/6}\label{fig.bblt}
\end{figure}

This profile, shown in figure \ref{fig.bblt}, only constitutes a weak solution to the equation (\ref{eq:generalporous}) as $\partial_x B$ is not continuous over all $x$ \citep{2013arXiv1312.0469H}. Due to this, a numerical scheme may misbehave around the boundaries, therefore for a convergence study it might be preferable to use a part of the profile far from the non-smooth points at $u=0$ for any numerical tests. This smooth region can be defined as 
\begin{equation}
    F(x,t)=\frac{1}{6t^{\frac{1}{3}}}\left(A - \frac{x^2}{t^{\frac{2}{3}}}\right).
\end{equation}
The boundaries in this solution are time dependant and the solution is only stable for $u>0$. 
There also exists a separable solution to the Boussinesq equation. This solution takes the form
\begin{equation}
    f(x,t)=\frac{-x^2}{6t}.
\end{equation}
The separable solution has the advantage of being smooth and continuous everywhere except at $t=0$, but does have time dependent boundary conditions when solved numerically. Additionally, this solution is not stable for any $t>0$, as the diffusion coefficient in equation (\ref{eq:porusmedium}) becomes negative, making it behave as an anti-diffusion equation. 
There is a further issue with all these solutions, that being that it is not useful for testing spatial convergence in any differencing scheme higher than the first order. Taking the case of a central difference scheme the spatial derivative can be written as, 
\begin{equation}
    \partial_x f_i=\frac{f_{i+1/2}-f_{i-1/2}}{h} + \mathcal{O}(h^2)
\end{equation}
with error terms depending on powers $h^2$ and higher. These error terms however also depend on the third spatial derivative of $f_i$ which is $0$ for this function, meaning that, at least at the first step, there is no spatial discretisation error. It can further be shown that provided the analytic solution can be discretised exactly in space, then the error at all times will not depend on the grid spacing (as shown in appendix \ref{appendix:errors}). This kind of problem is quite useful as a test of the time marching scheme, this is because no error arises from the spatial discretisation.

\section{Problematic shocks}\label{sec:Prob}
When studying the solutions to terminal velocity approximation and two fluid shocks, we found some discrepancies in the solutions. Already from a first analysis one can see that in the terminal velocity approximation solution is symmetric and smooth. This does not resemble the two fluid model shocks, where the shock front  precedes a region where the velocity difference between the gas and dust decays. 
\subsection{Two fluid shocks}
Solutions describing a two fluid shock can be found, this is not limited to perfectly coupled cases like the ones shown by \cite{LaibePriceDust}. For initial conditions with a jump in a gas there exists a well studied solution in the form of the Sod shock tube \citep{SOD19781}, such a solution has not been found for two fluid approximation dusty gases. It is however possible to find steady state solutions for shock fronts in this kind of setup \citep{tw0_fluid_shocks_wardle}. Starting from the two fluid equations for an isothermal gas coupled with a pressureless gas equation (\ref{eq:twofluid.1}) expressed in one dimension,
\begin{align}
&\partial_t \rho_g +\partial_x (\rho_g u_g)  =0 \label{eq:two_fluid_one_d.1}\\
&\partial_t( \rho_g u_g) +\partial_x(\rho_g u_g^2 + c_s^2 \rho_g)  = \frac{-\rho_d \rho_g}{t_s \rho}(u_g - u_d) \label{eq:two_fluid_one_d.2}\\
&\partial_t \rho_d +\partial_x(\rho_d u_d)  = 0 \label{eq:two_fluid_one_d.3}\\
&\partial_t(\rho_d u_d) + \partial_x(\rho_d u^2_d)  = -\frac{\rho_d \rho_g}{t_s\rho}(u_d-u_g).\label{eq:two_fluid_one_d.4} 
\end{align}
Considering a small region around the shock such that the source terms become negligible, we have conserved quantities
\begin{align}
B_g&=\rho_g u_g\label{eq:massReimann},\\
D_g&=\rho_g(u_g^2+c_s^2)\label{eq:momentumReimann},
\end{align}
which remain constant over the shock. Assuming that there is no jumps in $\rho_d$ and $u_d$ the mass fluxes $B_g$ and $B_d$ must be constant. For a steady state shock 
\begin{align}
&\partial_x(\rho_g u_g^2 + c_s^2 \rho_g)  = \frac{-\rho_d \rho_g}{t_s \rho}(u_g - u_d) \label{eq:two_Stat_shock.1}\\
&\partial_x(\rho_d u^2_d)  = -\frac{\rho_d \rho_g}{t_s\rho}(u_d-u_g).\label{eq:two_Stat_shock.2}
\end{align}
Considering these invariants it is possible to construct a first order differential equation to describe the profile of the shocks. The momentum equation can be rewritten as an ODE
\begin{equation}
\frac{d}{dx}\left(\frac{-B_g^2+c_s^2 \rho_g^2}{\rho_g^2}\right)=-\frac{1}{t_s\rho}\left(\rho_d B_g - \rho_g B_d\right),
\end{equation}
and by adding equations (\ref{eq:two_Stat_shock.1}) and (\ref{eq:two_Stat_shock.2}) it is possible to eliminate $\rho_d$ as 
\begin{equation}
    \rho_d=\frac{B^2_d}{D - B_g^2/\rho_g - c_s^2\rho_g},
\end{equation}
where $D=\rho_g(u_g^2 + c_s^2)+\rho_d u_d^2$ must be constant. This operation yields the ODE 
\begin{equation}
    \label{eq:twoFluidODE}
    \frac{d\rho_g}{dx} \left(\frac{-B_g^2+ \rho_g^2 c_s^2}{\rho_g^2}\right)=-\frac{B_d}{t_s}\left(\frac{B_d B_g-D\rho_g+B^2_g +c_s^2\rho^2_g}{D\rho_g - B_g^2-c_s^2\rho_g^2+ B^2_d}\right),
\end{equation}
which can be solved numerically or analytically to produce a density and thus a velocity profile for both the dust and gas. Studying the solution and experiments, it can be seen that when the preshock velocity $u_R>c_s$ then for a length scale $L_{\mathrm{equil}}\sim c_s t_s$ the velocity difference between the dust and gas $\Delta u \sim c_s$. 
\subsection{Shocks in the terminal velocity approximation}
While the terminal velocity approximation yields well-defined, smooth "shock" profiles (see section \ref{sec:Test:shock}), around shocks the approximation breaks down, rendering these solutions unphysical. The terminal velocity approximation relies on the stopping distance of the dust being very small compared to any other length scale in the fluid. Around a shock this ceases to be true, as the stopping length becomes very long compared to the length scale of the shock. By expanding the equations it is possible to show that in fact the approximation cannot hold around shocks. 
When describing stationary dusty shocks with a two fluid approximation, there is a finite length over which the gas velocity $u_g$ jumps for supersonic to subsonic, while the dust velocity is unchanged. Therefore just behind the shock there is a region where $\Delta u \sim c_s$. In the one fluid dusty shock the gas and dust are assumed to be coupled, therefore this region does not appear in this approximation. This results in a first order error around the shock with respect to the two fluid shock. In dimensionless form this error can be written as
\begin{equation}
\delta(\Delta u) \sim \frac{c_s^2 u t_s}{u^2 L} = \frac{\mathrm{St}}{\mathrm{Ma^2}}.
\end{equation}
This first order error means that around shocks the terminal velocity approximation breaks down. The only way to approach this problem within the terminal velocity approximation would be to resolve the shock, through higher shock viscosity. Alternatively one could under resolve the shock such that the region where the error appears is much smaller than a single grid cell. Neither of these solutions is fully satisfactory, as increasing the shock viscosity will produce unphysical results, while limiting the resolution might be unsuitable for some problems.

\section{Computational methods}
\subsection{A DUSTY implementation of FARGO3D}\label{sec:methodFargo}
In a diffusion solver for a hydrodynamics code there is an inherent tradeoff between stability and speed. Fast and accurate explicit solvers are unstable for long timesteps, as their stability condition is inversely proportional to the spatial resolution squared while the advection step's stability condition is dependant on just one over the spatial resolution. Implicit integrators allow for longer steps, as they are either unconditionally stable or have a much wider stability region than explicit integrators. Implicit integrators however are less accurate, more computationally expensive, and much more difficult to parallelise. A lot of progress has been made in the field of explicit stiff ODE solvers and these stable solvers can be used to do the timestepping in semi-discretised PDE integrators. We implement a stable integrator to integrate the diffusion term $C(P,\rho)=c_s^2\nabla\cdot\left[f_d\nabla P\right]$ in equation (\ref{eq:onefluiddim.3}).
This is the source term from the locally isothermal terminal velocity approximation derived in (\cite{LinYoudin2017}), see equation (\ref{eq:onefluiddim.1}-\ref{eq:onefluiddim.3}). Due to FARGO3D being operator split, it is possible to integrate the source terms separately from the advection terms, and produce a solution by adding the two. 
In our problem, we implemented a stable solver to integrate the diffusion term $C\left(P,\rho\right)$ in a finite difference discretised scheme. For this implementation, we built upon and modified the existing and well tested FARGO3D \citep{Fargo3d2016} operator split code, and added a separate solver to integrate the diffusion term $C(P,\rho)$; the Runge Kutta Legendre second order (RKL2) scheme derived by \cite{balsaraRKL}. This algorithm provides several benefits with respects to both explicit and implicit schemes. By sacrificing higher order, the scheme gains the stability of an implicit scheme, but at the same time maintains the clarity and ease of paralellisation of an explicit scheme. RKL is not the only unconditionally stable scheme, examples of other explicit unconditionally stable schemes being RKC and ROCK (constructed around Chebyshev polynomials). The stability of these schemes for the integration of parabolic PDEs was analysed in \cite{supertimestepping}, showing that these schemes, when correctly formulated are stable for any step. These RKC methods however have smaller stability regions than RKL methods, meaning that they can produce larger features not present in the true solution. The other implicit approaches are usually inefficient as they require large matrix inversions, as shown in \cite{supertimestepping} and \cite{1742-6596-837-1-012016}. This specific scheme (RKL2) was picked because it was shown to be well behaved with non linear diffusion problems by \cite{balsaraRKL}, as well as being a very efficient scheme (\cite{1742-6596-837-1-012016}). The RKL2 scheme maintains unconditional stability by being constructed on a Legendre polynomial, which when chosen correctly has a magnitude of $1$, therefore there can be no growing wavemodes, making the scheme stable. For our FARGO3D implementation we used a central finite difference discretisation in space with the previously described RKL2 integrator for the timestepping. This implementation plugs into the FARGO3D machinery without any major changes to the core solver. Some change had to be made to the calculation of the Courant number, or CFL, as the soundspeed in the dusty gas is calculated differently from either an isothermal single gas or an adiabatic gas. We took the gas only sound speed to be the sound speed, as the dust gas sound speed is guaranteed to be smaller than the pure gas sound speed, therefore the CFL step cannot be overestimated, and the code remains stable, at the expense of a slightly shorter timestep than could be possible.

\section{Results}\label{sec:Res}
\begin{figure}
    \includegraphics[width=\columnwidth]{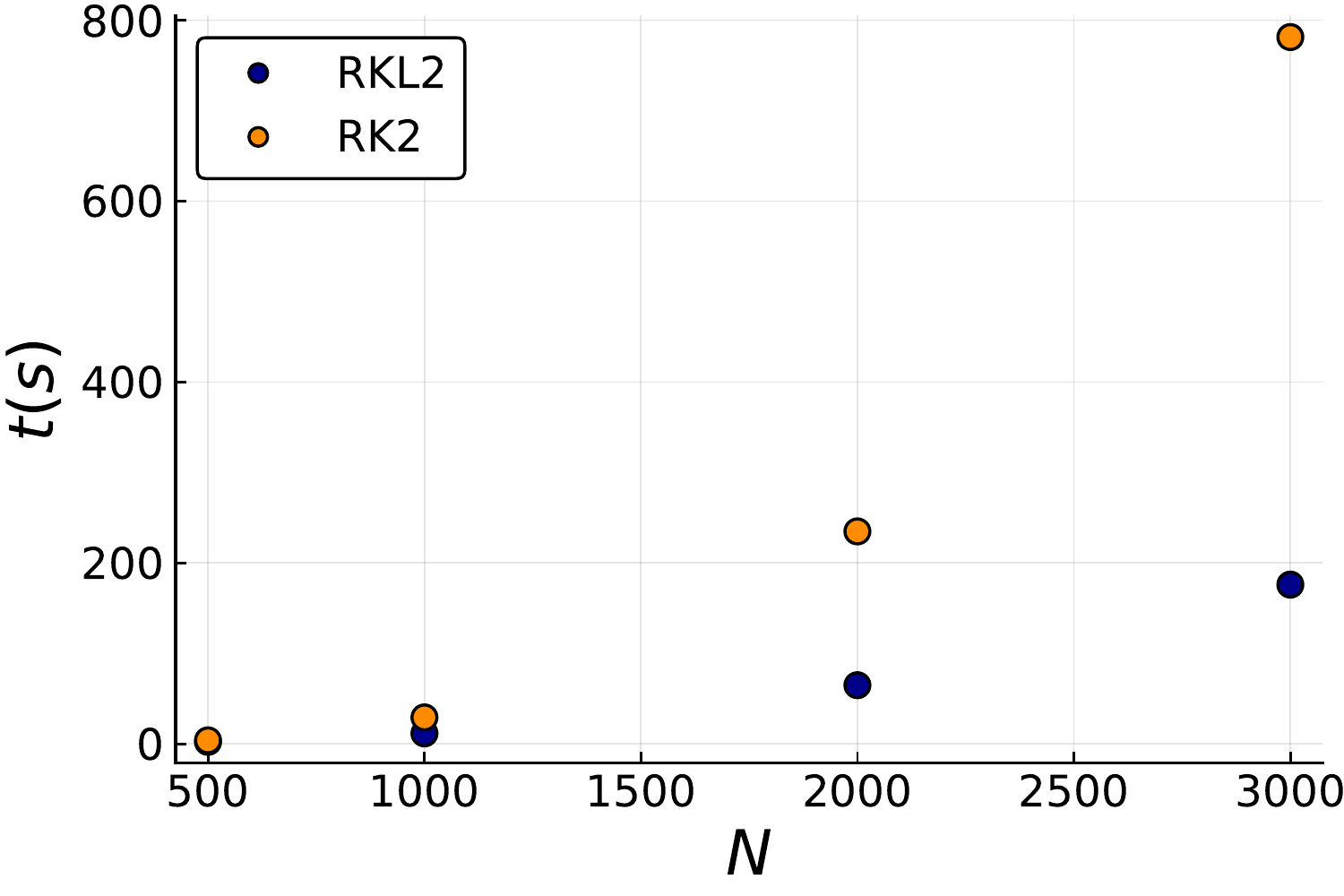}
    \caption{Comparison of time taken for RK2, and RKL2 schemes to run until $t=10$ in the wave test, using the same parameters as the wave test.}
    \label{fig:Timing}
\end{figure}

\begin{figure}
    \subfloat[RK2 \label{sfig:RK2Conv}]{
        \includegraphics[width=\columnwidth]{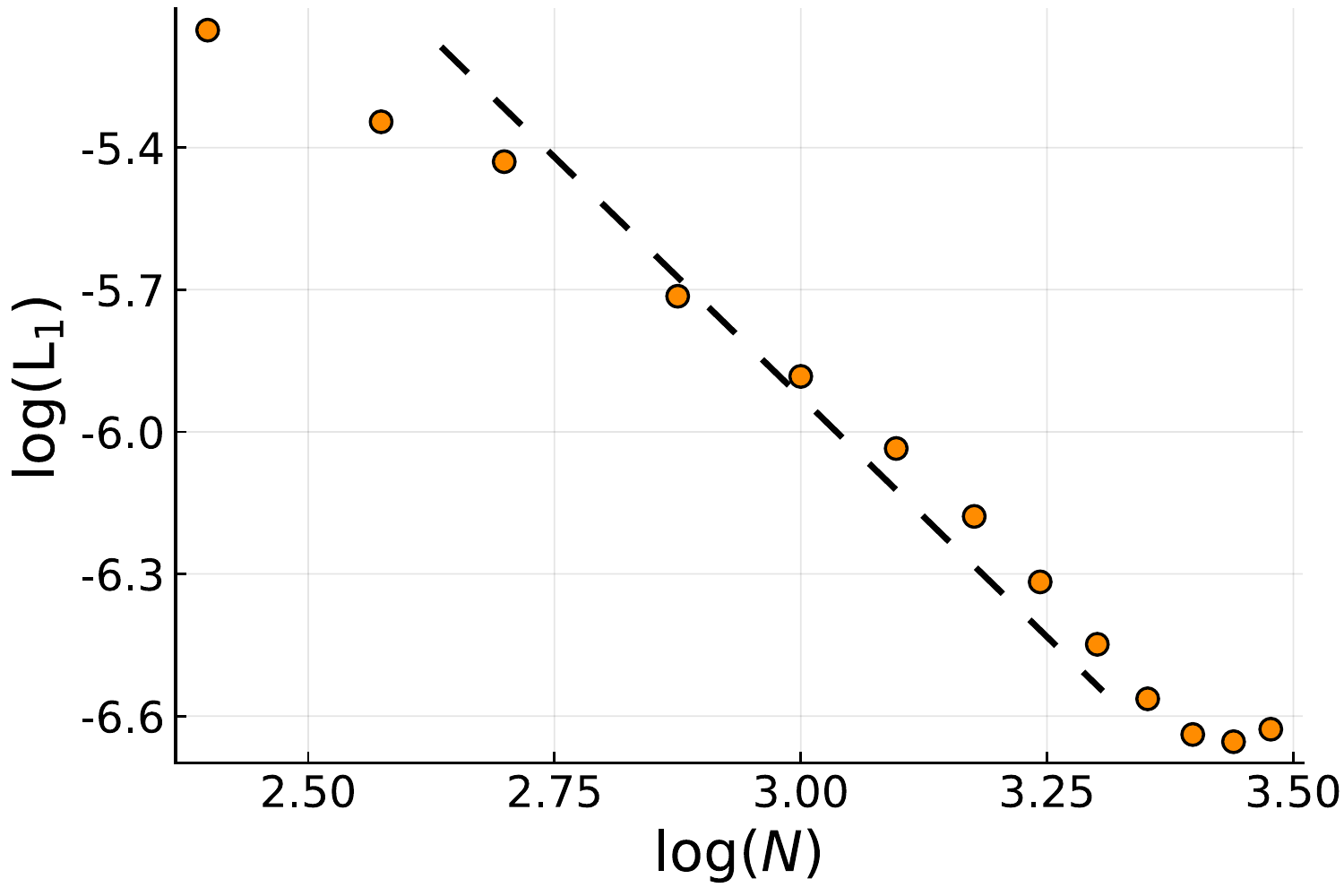}
    }
    
    \subfloat[RKL2 \label{sfig:RKL2ConvStable}]{
        \includegraphics[width=\columnwidth]{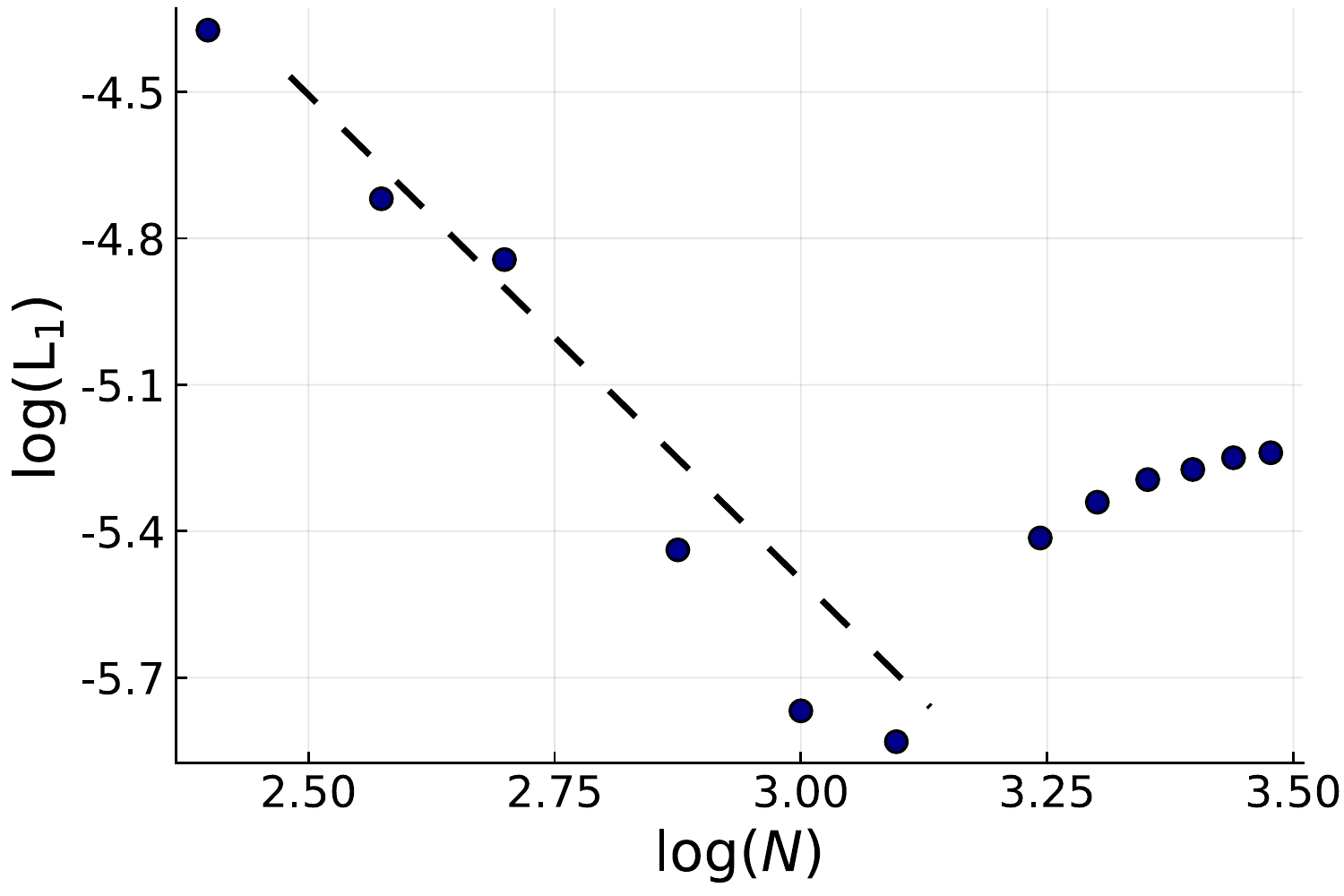}
    }
    \caption{Convergence test of wave propagation problem, using two explicit timestepping schemes, RK2 (a) and unconditionally stable RKL2 (b). The error is measured using an L1 norm, while N is the number of cells for a fixed box size. The dashed line shows second order convergence.}
    \label{fig:Convergence}
\end{figure}

We used the solutions from section \ref{sec:Tests} to test the performance of our dusty version of FARGO3D described in section \ref{sec:methodFargo}. The dusty wave solution was used to run a convergence study for the code. For these runs we used a box size of 10.0 with a wavenumber $k=2\pi/10$. The background values used were $ \rho_0 =50.0$, $P_0=25.0$, $v_0=0$, $c_s=1.0$ and $t_s=1.0$. The perturbation was given by taking a density perturbation amplitude of $\rho_1=0.01$, all the other perturbation amplitudes being constrained by setting one. The code was run until $t=10.0$, then an $L_1$ norm was taken to evaluate the numerical error of the scheme. The RKL2 implementation of the code performed well showing comparable spatial convergence to an RK2 scheme (Fig. \ref{fig:Convergence}). The absolute error in the RKL2 scheme is larger than that exhibited by the RK2 scheme, however the RKL2 scheme is much faster than its fully explicit counterpart for any high resolution problem, as the number of calculations scales with $\frac{1}{h^2}$ not $\frac{1}{h^3}$ as for classic explicit schemes (like RK2), this difference can be clearly seen in fig. \ref{fig:Timing}. 

\begin{figure}
    \subfloat[RK2 \label{sfig:RK2Conv1}]{
        \includegraphics[width=\columnwidth]{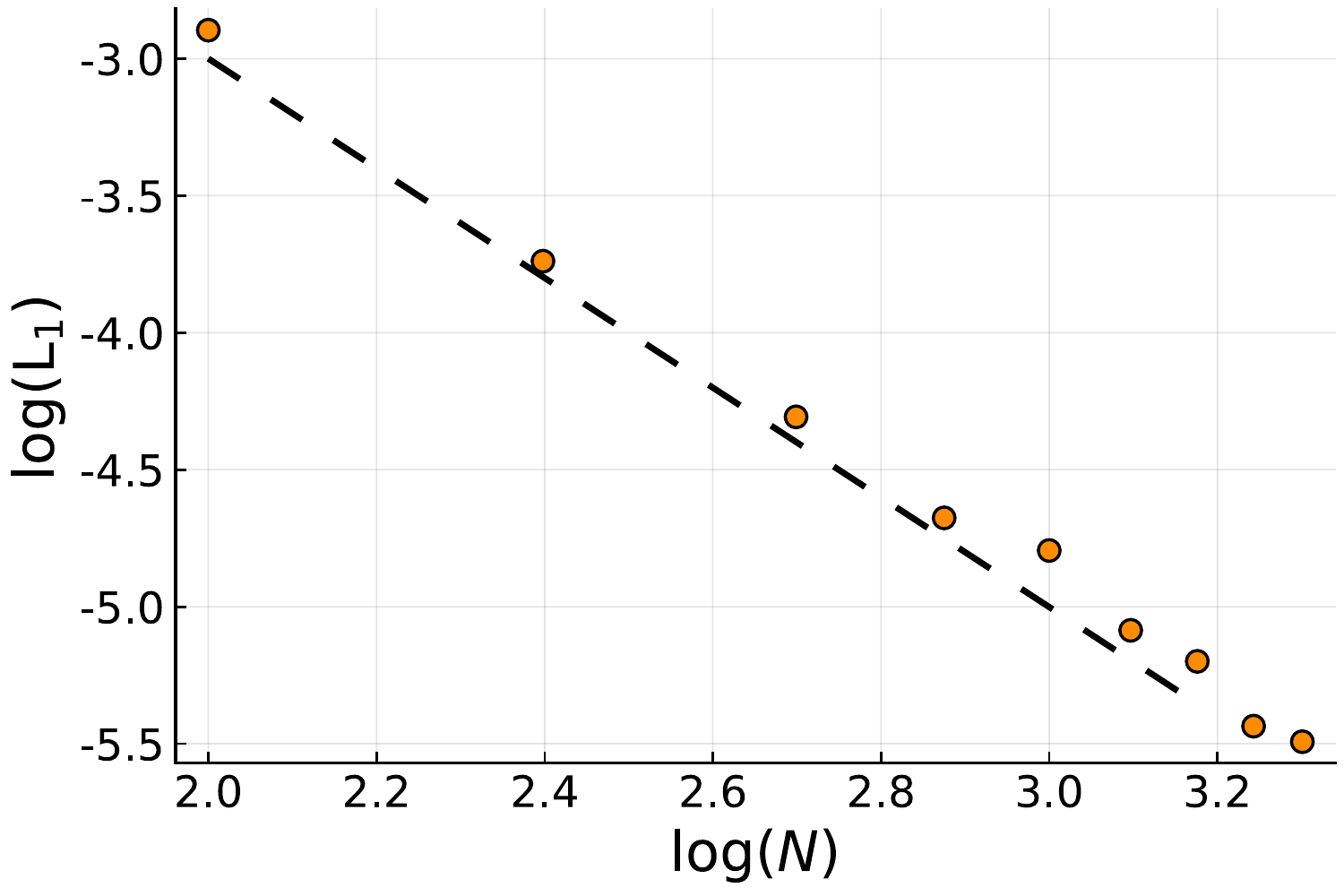}
    }
    
    \subfloat[RKL2 \label{sfig:RKL2ConvStable1}]{
        \includegraphics[width=\columnwidth]{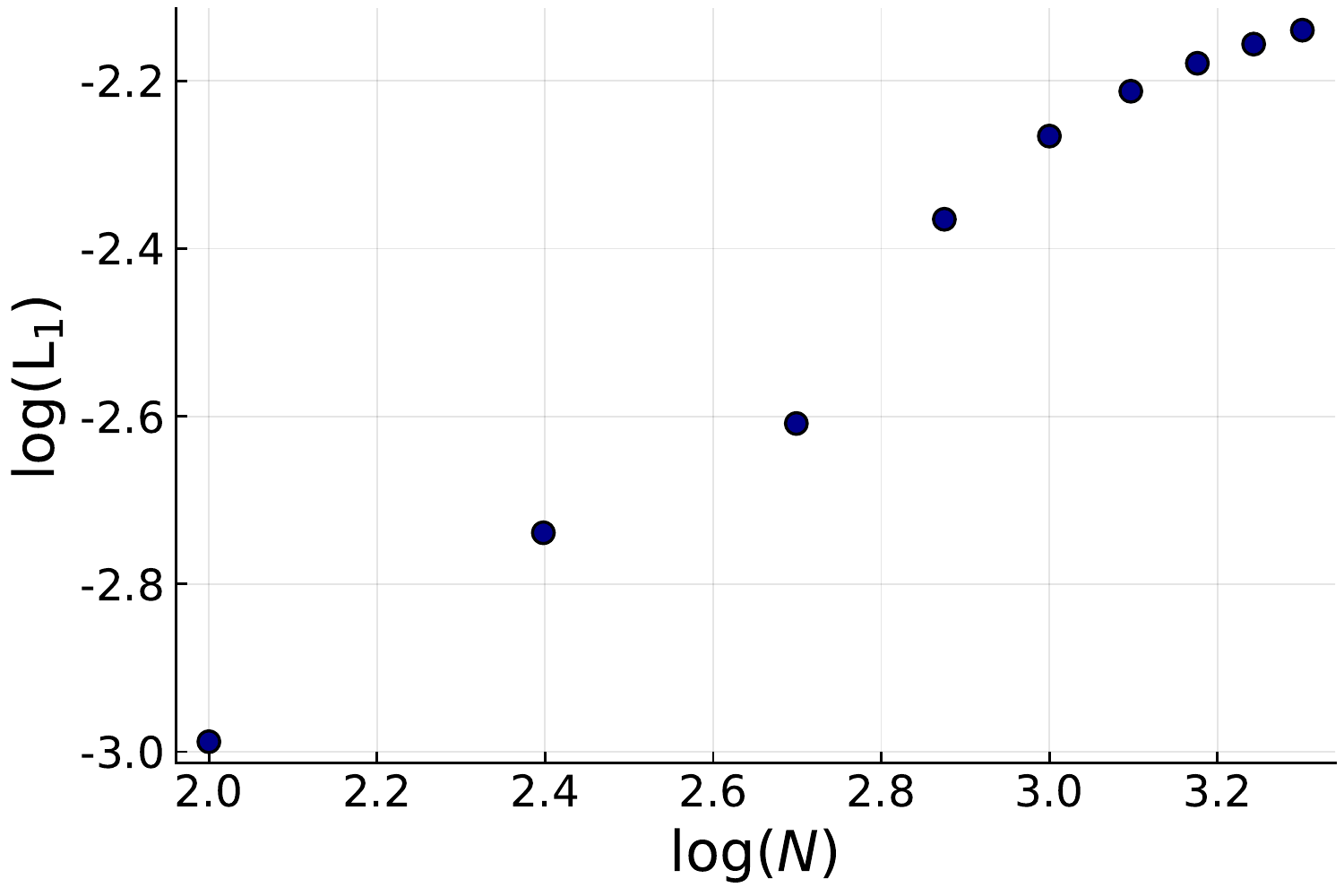}
    }
    \caption{Convergence test on the diffusion problem from section \ref{sec:Tests}, using the two explicit timestepping schemes, RK2 (a) RKL2 (b). The error is measured using an L1 norm, while N is the number of cells for a fixed box size. The dashed line shows second order convergence.}
    \label{fig:Convergence1}
\end{figure}

For the diffusion test we ran the diffusion problem from section \ref{sec:Tests}, using the central portion of a Barenblatt solution with $A=36$ in the region $-1<x<1$, far from the non-smooth points. We ran the problem for a time of 0.1 from $t=1.0$ to $t=1.1$. To run this problem we turned off the transport portion of FARGO3D and allowed the code to take the maximum stable time-step (RKL2 being unconditionally stable only ever takes 1 step with a varying number $s$ of supersteps). This experiment shows that increasing the number of RKL stages does not increase the precision of the scheme, only its stability as can be seen in figure \ref{fig:Convergence1}. In fact the accuracy of an $s$ stage RKL scheme slightly decreases with increasing $s$. This becomes especially obvious in the diffusion test, where we advance only the diffusion term in time. By increasing the spatial resolution in this test we do not reduce the spatial error, but we force smaller timesteps in the RK2 scheme to maintain stability, increasing the precision. In the RKL2 scheme however, the timestep is not reduced, as the number of supersteps is increased to maintain stability instead, leading to the error linearly increasing with the number of supersteps. This in turn means that the RKL2 method is faster, especially at high resolution, being about four times as fast for the $N=3000$ test in terms of total run time (fig. \ref{fig:Timing}).

\begin{figure}
    \includegraphics[width=\columnwidth]{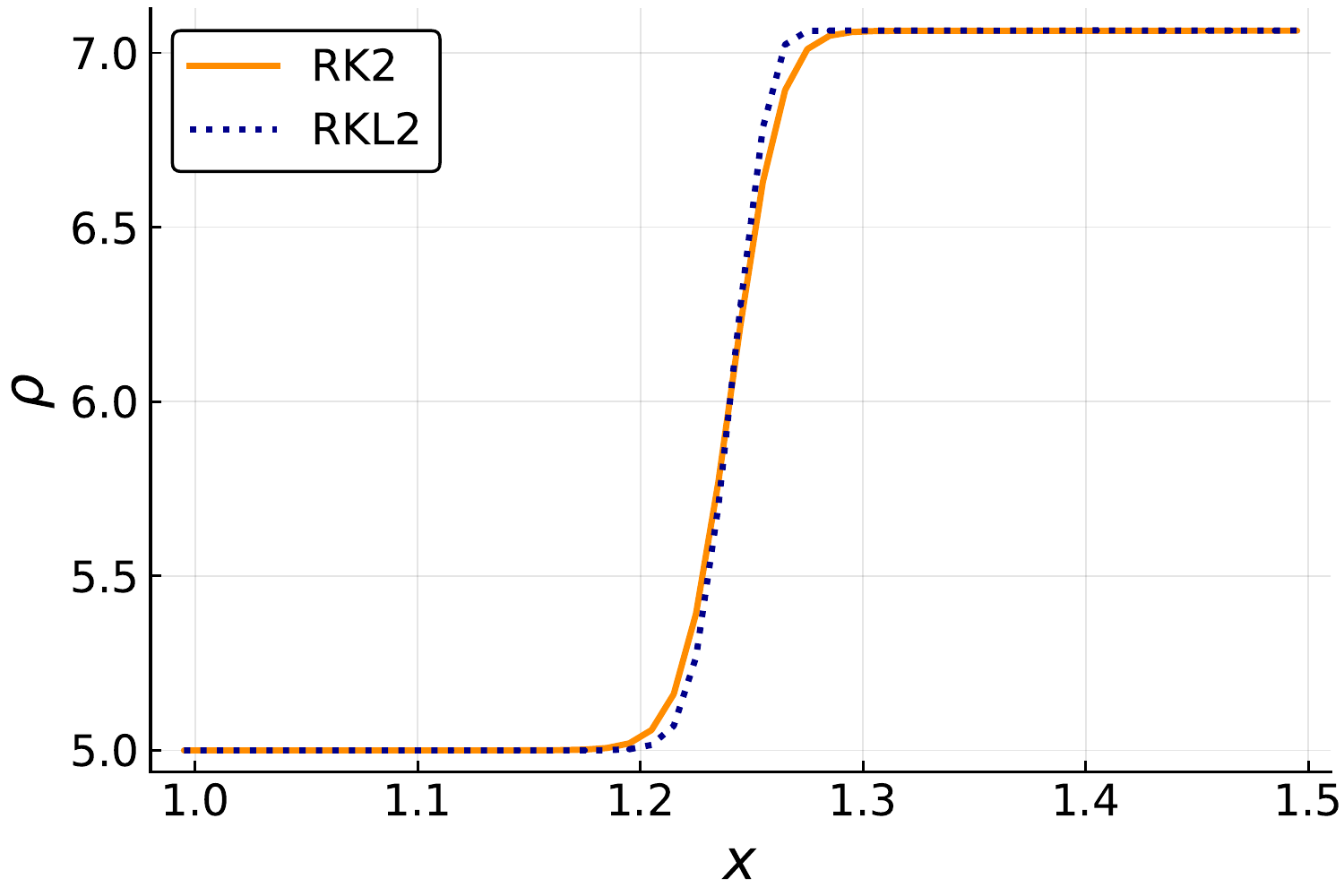}
    \caption{Comparison of a shock profile using the two explicit timestepping schemes, RK2, and RKL2.}
    \label{fig:ShockNum}
\end{figure}

Regarding the stability of the scheme as a whole, the code is able to remain stable running a discontinuous initial condition, with the shock produced converging on the steady state shock profile calculated in section \ref{sec:Test:shock}, as expected. For this test we used a resolution of 100 cells per unit length, an initial density and pressure jump of a factor of 2 from $\rho_R=5.0$ to $\rho_L=10.0$ the other parameters being: $t_s=0.1$, $c_s=1.0$, and $f_d=0.5$. The RKL2 scheme under diffused the solution compared to the RK2 scheme as can be seen in figure \ref{fig:ShockNum}. No real statement can be made here about the relative accuracy of the schemes, as we do not have a time dependent solution for the shock wave. The analytic solution holds as $t\rightarrow\infty$ and we expect the time dependent solution to tend to our analytic solution in the large t limit, but we do not know how a discontinuity evolves towards said solution.

\section{Discussion}\label{sec:Dis}
\subsection{Dusty shocks}
For an astrophysical flow model it can be argued that the inability to describe shocks is an important flaw. Most astrophysical flows are largely very low density and dominated by external fields, this means that supersonic flows and shocks are quite common. The fact that shocks are common does not mean that there are no scenarios where shocks are not important. In protoplanetary discs, for example the dynamics of many phenomena like the evolution of vortices, and the linear growth of certain instabilities like the streaming instability, or the evolution of the dust distribution in the disc can be studied without needing to model shocks. In these cases the pressure profile is smooth and shocks are not expected to arise.

As we have shown the terminal velocity approximation breaks down around shocks. This means that the local behaviour around shocks is not to be trusted in terminal velocity approximations. If the shock is important to the evolution of the system, one should not expect to obtain correct results from a terminal velocity approximation. In essence the terminal velocity approximation is not well suited for studying structure arising because of or at shocks.

It is not clear however to what extent the error around shocks is an issue for the global behaviour of the system. If the shocks are short lived, and only affect a small area of the domain, the final result may be largely unaffected by the error appearing due to the shocks. Additionally, the length scale of the error is quite small, so it might not be a very serious issue if a shock does arise. As the length scale the error occurs on may, but is not guaranteed to, be much smaller than the grid scale; making the terminal velocity approximation nearly indistinguishable from the two fluid solution. This does not change the fact that caution should be taken when shocks arise in a terminal velocity approximation simulation.

With regards to protoplanetary discs, shocks are probably important in the evolution of the dust distribution, as well as the global evolution, contributing to among other things, the vertical mixing in the disc \citep{0004-637X-641-1-534} and gaps in the dust density. This limits the useful scope for the terminal velocity approximations, as the main advantages of one fluid approximations over two fluid models are the reduced computational cost and better scaling with system size, making terminal velocity approximations very well suited for high resolution global simulations of discs. The inability to model shocks  means one has to be careful when using the terminal velocity approximation in cases where shocks may be important.
\subsection{Slightly viscous dusty gases}
Evidence suggests that while protoplanetary discs probably have a very low turbulent viscosity, the value of this viscosity is probably non zero \citep{Flaherty_2017}. Having a model to simulate slightly viscous dusty gases is therefore of great value. The result obtained in section \ref{sec:Visc} is especially relevant in that the dusty viscous stress tensor is identical to a pure gas stress tensor and would thus require no modification to a code to be implemented. The addition of a viscosity model is also important in that it adds a way to smooth shocks. As discussed in section \ref{sec:Prob} if the velocity jump in the shock is very small, then the first order error does not appear. In a more viscous discs very strong shocks are less common, this increases the range of problems to which the terminal velocity approximation could be safely applied to. 

\subsection{Dusty vortices}
We propose that this kind of model is very well suited for the simulation of dusty vortices in protoplanetary discs. To fully model the stability of vortices, a high resolution 3D simulation is required, which is made feasible by the use of a terminal velocity approximation. If the problem is then chosen correctly, i.e. the vortex radius is smaller than the local disc scale height \citep{2010ApJ...725..146P, 2001ApJ...552..793G}, then no shocks are expected to arise. This kind of setup can be used to investigate the potential of vortices to act as dust traps to facilitate grain growth, or the long term stability of vortices in dusty discs. As mentioned earlier shocks are not present in this setup, so a very non dissipative scheme could also be used, for example a spectral scheme, which allows for the viscosity to be controlled and minimised. This is especially useful in the case of an inviscid disc, where the vortex lifetime would be controlled by the dust density alone.

\section{Conclusions}\label{sec:Conc}
Single fluid terminal velocity approximations for dusty gases are very promising techniques for the study the evolution of a dusty gas in a protoplanetary disc. The computational complexity of this kind of approximation is lower than in a two fluid approach, especially as the correct dust velocity is recovered even in the long timestep limit on one fluid approximations, which is not the case for two fluid solvers.  
In this paper we show that the approximation is not limited to inviscid fluids, as for an $\alpha$ viscosity model the viscous stress tensor is unchanged at first order in Stokes number. This result implies that there is no need to modify the one fluid description for it to be compatible with $\alpha$ viscosity model, by consequence it is simple to implement the terminal velocity approximation in an $\alpha$ viscosity model, as no additional changes are required in comparison to an inviscid solver. 
We find however an important limitation in the terminal velocity approximation, being its inability to model shocks correctly. Around the shock an error of order the soundspeed is made in calculating the barycenter velocity, and the correct structure of the shock is not retrieved.  The scale of the error is rather small, given by $\mathrm{Ma}_{\mathrm{shock}} c_s t_s$, the product of stopping time sound speed and shock Mach number, this can however be larger than the gridscale in a high resolution simulation or in cases where $\mathrm{Ma}$ is very large, so care must be taken when simulating initial conditions that may see shocks arising. While the terminal velocity approximation is not suitable for studying shocks in protoplanetary discs, however there are many cases in planet formation and evolution where this is not an issue. A good example of this is small (less than 1 scale height in radius) vortices. These, in a protoplanetary disc are very important to the evolution of dust distribution, while not giving rise to shocks. Because of this, the terminal velocity approximation is especially well suited to study the behaviour of dust laden vortices. This will be the subject of the next paper in this series.

\section*{Acknowledgements}
We thank the referee, Jeff Oishi, for his thorough response, which much improved the quality and clarity of our paper.
FL is funded by an STFC studentship. SJP is funded by a Royal Society University Research Fellowship.



\bibliographystyle{mnras}
\bibliography{LovascioPaardekooper} 




\appendix

\section{Proof of non mixing errors for quadratic in space problems solved with centered differences schemes}\label{appendix:errors}
Taking a numerical approximation $\Tilde{u}_i^n$ of an analytic function $u_i^n$ constructed at timestep $n$ at grid node $i$ using a centered in space and $A$th order explicit in time scheme. One can write this as,
\begin{equation}
    \Tilde{u}_i^n=u_i^n + \epsilon_i^n\left(h^2,\tau^A\right) 
\end{equation}
where $\epsilon_i^n$ is the numerical error at timestep $n$ in cell $i$. The error term $\epsilon_i^n$ can be expanded such that,
\begin{align}
    \epsilon_i^n\left(h^2,\tau^A\right)&=\partial_x^3(\Tilde{u}_i^{n-1})\mathcal{O}(h^2)+\partial_t^A(\Tilde{u}_i^{n-1})\mathcal{O}(\tau^A) \\
    &=\partial_x^3(u_i^{n-1}+\epsilon_i^{n-1})\mathcal{O}(h^2)+\partial_t^A(u_i^{n-1}+\epsilon_i^{n-1})\mathcal{O}(\tau^A).
\end{align}
Postulating that
\begin{equation}\label{app:Postulate}
    \Tilde{u}_i^n=u_i^n + \epsilon_i^n\left(\tau^A\right), 
\end{equation}
meaning that the numerical error at any time depends only the time step and not on the spatial step. One can show that given that \ref{app:Postulate} is true for $n$ then it is also true for $n+1$. Starting with the general expression for the numerical solution at $n+1$,
\begin{equation}
    \Tilde{u}_i^{n+1}=u_i^{n+1} + \epsilon_i^{n+1}\left(h^2,\tau^A\right),
\end{equation}
the error terms then become,
\begin{align}
    \epsilon_i^{n+1}\left(h^2,\tau^A\right)&=\partial_x^3\left(\Tilde{u}_i^{n}\right)\mathcal{O}\left(h^2\right)+\partial_t^A\left(\Tilde{u}_i^{n}\right)\mathcal{O}\left(\tau^A\right) \\
    & \begin{aligned}
        &=\partial_x^3\left[u_i^{n}+\epsilon_i^{n}\left(h^2,\tau^A\right)\right]\mathcal{O}\left(h^2\right)\\
        &+\partial_t^A\left[u_i^{n}+\epsilon_i^{n}\left(h^2,\tau^A\right)\right]\mathcal{O}\left(\tau^A\right).
    \end{aligned}
\end{align}
From \ref{app:Postulate} it can be seen that, 
\begin{equation}
    \epsilon_i^{n}\left(h^2,\tau^A\right)=\epsilon_i^n\left(\tau^A\right)
\end{equation}
therefore,
\begin{equation}
    \partial_x^3\left[u_i^{n}+\epsilon_i^{n}\left(h^2,\tau^A\right)\right]\mathcal{O}\left(h^2\right)=\partial_x^3 u_i^{n}\mathcal{O}\left(h^2\right).
\end{equation}
From our initial assumptions about the analytic function however we know that it can only be quadratic in space all times for which it is well posed, making the third spatial derivative $0$. The error at step $n+1$ can then be rewritten as 
\begin{equation}
    \epsilon_i^{n+1}\left(h^2,\tau^A\right)=\epsilon_i^{n+1}\left(\tau^A\right).
\end{equation}
Given that at $n=1$ \ref{app:Postulate} is trivially true, as $\Tilde{u}_i^0\equiv u_i^0$ due to $\Tilde{u}_i^0$ being the initial conditions, then by induction, a numerical approximation $\Tilde{u}_i^n$ to a solution $u_i^n$, analytically discretisable at any time, must obey the identity
\begin{equation}\label{app:proved}
    \Tilde{u}_i^n=u_i^n + \epsilon_i^n\left(\tau^A\right) \quad \forall \quad n \in \mathbb{N} 
\end{equation}


\label{lastpage}
\end{document}